\documentclass[a4paper]{spie}  %a4paper>>> use this instead for A4 paper

\usepackage[utf8]{inputenc} 
\usepackage[T1]{fontenc}
\usepackage{amsmath,amsfonts,amssymb}
\usepackage{graphicx, booktabs, multirow}
\usepackage[colorlinks=true, allcolors=black]{hyperref}
\usepackage[nolist]{acronym}
\usepackage{siunitx}
\usepackage{nicefrac}
\usepackage{float, enumerate, adjustbox}
\usepackage{stackengine}

\title{An open-source software platform for translational photoacoustic research and its application to motion-corrected blood oxygenation estimation}

\author[a,b]{Thomas Kirchner\,}
\author[a,c]{Janek Gröhl\,}
\author[a,b]{Franz Sattler\,}
\author[d]{Moritz S. Bischoff\,}
\author[a]{Angelika Laha\,}
\author[e,c]{Marco Nolden\,}
\author[a,c]{Lena Maier-Hein\,}

\affil[a]{ Division of Computer Assisted Medical Interventions (\textsc{CAMI}), German Cancer Research Center (\textsc{DKFZ}), Heidelberg, Germany}
\affil[b]{ Faculty of Physics and Astronomy, Heidelberg University, Germany}
\affil[c]{ Medical Faculty, Heidelberg University, Germany}
\affil[d]{ Department of Vascular and Endovascular Surgery, Heidelberg University Hospital, Heidelberg, Germany}
\affil[e]{ Division of Medical Image Computing (\textsc{MIC}), German Cancer Research Center (\textsc{DKFZ}), Heidelberg, Germany}

\authorinfo{Please address your correspondence to Thomas Kirchner, e-mail: t.kirchner@dkfz-heidelberg.de}

\begin{document} 
\maketitle 

\begin{abstract}
Photoacoustic (PA) imaging systems based on clinical linear ultrasound arrays have become increasingly popular in translational PA research. Such systems can be more easily integrated in a clinical workflow due to the simultaneous access to ultrasonic imaging and their familiarity of use to clinicians. In contrast to more complex setups, hand held linear probes can be applied to a large variety of clinical use cases. However, most translational work with such scanners is based on proprietary development and as such not accessible to the community.
In this contribution, we present a custom-built, hybrid, multispectral, real-time photoacoustic and ultrasonic imaging system with a linear array probe that is controlled by software developed within the Medical Imaging Interaction Toolkit (MITK) a highly customizable and extendable open-source software platform. Our software offers direct control of both the laser and the ultrasonic system and may serve as a starting point for various translational research projects and developments. To demonstrate the applicability of the platform, we used it to implement a new method for blood oxygenation estimation in the presence of non-rigid inter-frame motion caused by pulsing arteries. Initial results from experiments with healthy human volunteers demonstrate the suitability of the method with the sample clinical application of imaging the common carotid artery as well as peripheral extremity vessels.

\end{abstract}

%>>>> Include a list of keywords after the abstract 


\begin{acronym}[TDMA]
	\acro{PA}{photoacoustic}
	\acro{DAS}{Delay and Sum}
	\acro{DMAS}{Delay Multiply and Sum}
	\acro{MV}{Minimum Variance}
	\acro{US}{ultrasonic}
	\acro{MITK}{Medical Imaging Interaction Toolkit}
	\acro{SNR}{signal-to-noise ratio}
	\acro{CNR}{contrast-to-noise ratio}
	\acro{GUI}{graphical user interface}
	\acro{API}{application programming interface}
	\acro{ITK}{Insight Toolkit}
	\acro{DAQ}{data acquisition}
	\acro{PAUS}{photoacoustic-ultrasonic}
\end{acronym}

\section{Introduction}
\label{sec:intro}
Estimating blood oxygenation ($\mathrm{sO_2}$) or similar functional parameters deep inside tissue is one of the key use cases for multispectral \ac{PA} imaging\cite{Taruttis2015}. A reliable and fast measurement of $\mathrm{sO_2}$ can have many applications and is one of the main arguments for translation of PA imaging into the clinic\cite{Gerling2014, Mohajerani2015}.
\ac{PA} imaging systems that are based on clinical ultrasound systems with hand held linear probes\cite{Kim2016} are relatively easily integrated in a clinical workflow due to the simultaneous access to ultrasonic imaging and the familiarity of use to clinicians. In contrast to other \ac{PA} imaging setups, hand held linear probes can be applied to a large variety of clinical use cases. However, most translational work with such scanners is based on proprietary software development and as such not accessible to the community.

To address this bottleneck, we present a real-time multispectral hybrid \ac{PA} and \ac{US} (PAUS) imaging system running on the open-source platform \acf{MITK}\cite{Nolden2013}. The functionality of the software framework is demonstrated by implementing a new method for $\mathrm{sO_2}$ estimation in the presence of motion. The latter relies on the computer vision method \emph{optical flow}\cite{Farneback2003} to co-register a sequence of \ac{PA} images using brightness patterns in corresponding \ac{US} images and was specifically designed to mitigate non-rigid inter-frame motion caused by pulsing arteries. The performance of the platform including the presented software modules is demonstrated by means of motion-corrected blood oxygenation estimation in the carotid artery and accompanying vein based on data acquired with the presented system. 

\section{PAUS imaging platform}
In this section, we present the hardware setup and the open-source software components developed for our PAUS imaging platform.

\begin{figure}[h!bt]
  \begin{center}
  \includegraphics{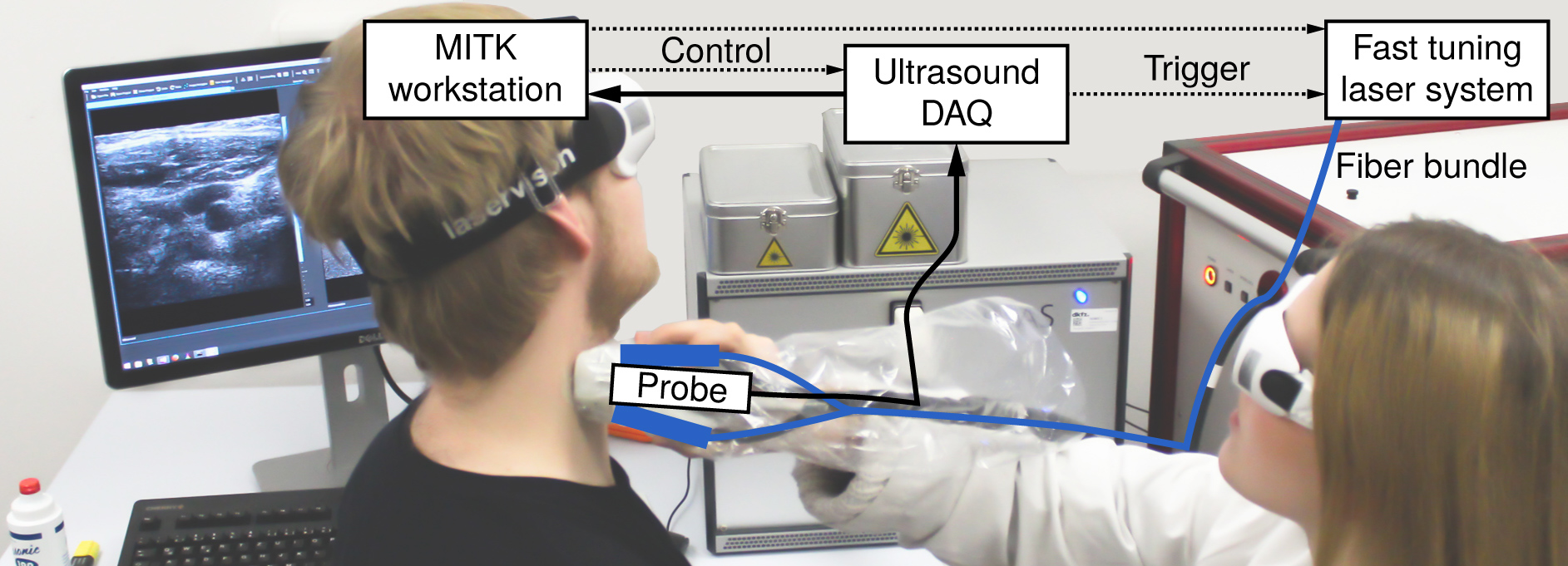}
  \end{center}
  \caption[Setup]
  {Hardware setup in use. The custom build \acf{PAUS} imaging system consists of a fast tuning optical parametric oscillator (OPO) laser triggered by the \acf{DAQ} system, which uses a linear transducer. The  workstation runs the \acf{MITK} for control of laser and \ac{DAQ} as well as real-time processing and visualization of the \acf{PA} and \acf{US} data streams.}
  \label{fig:hardware} 
\end{figure}

\subsection{Hardware setup}
\label{sec:hw}
The hardware setup is shown schematically in Fig.\,\ref{fig:hardware}. Data acquisition is performed using our DiPhAS US system (Fraunhofer IBMT, St.~Ingbert, Germany) and a 128-element linear US transducer operating on a center frequency of 7.5\,MHz (L7-Xtech, Vermon, Tours, France). The light from a fast tuning OPO laser (Phocus Mobile, Opotek, Carlsbad, USA) is delivered via a custom fiber bundle to the probe and into the tissue or phantom. 
The pulse energy of up to 40\,mJ at a pulse repetition rate of 20\,Hz is delivered over a surface of 2\,cm$^2$, below the maximum permissible exposure safety limit \cite{ANSI2005} to avoid tissue damage. The actual energy output of each individual laser pulse can be measured with an integrated pulse energy sensor. 
For all mentioned hardware components application programming interfaces (APIs) are provided by the vendors. Additionally, several software components were specifically developed by us for hardware communication, as detailed below.

\subsection{MITK Photoacoustics software components}
\label{sec:sw}
The real-time PAUS imaging software has been developed as part of \ac{MITK}, a free open-source software platform for interactive medical image processing software. MITK has a modular architecture, where \emph{Plugins} contain the user interface and have dependencies to \emph{Modules} which comprise the application logic and domain specific functionality. Each plugin provides graphical user interfaces (GUIs) for specialized controls in so-called \emph{Views} which are callable from the MITK workbench -- the interactive user application for research within MITK.
The structure of MITK therefore implements a clear separation of the application layer, i.e.\ the direct user interaction, and lower layers such as hardware control. 

\begin{figure}[h!bt]
  \begin{center}
  \includegraphics{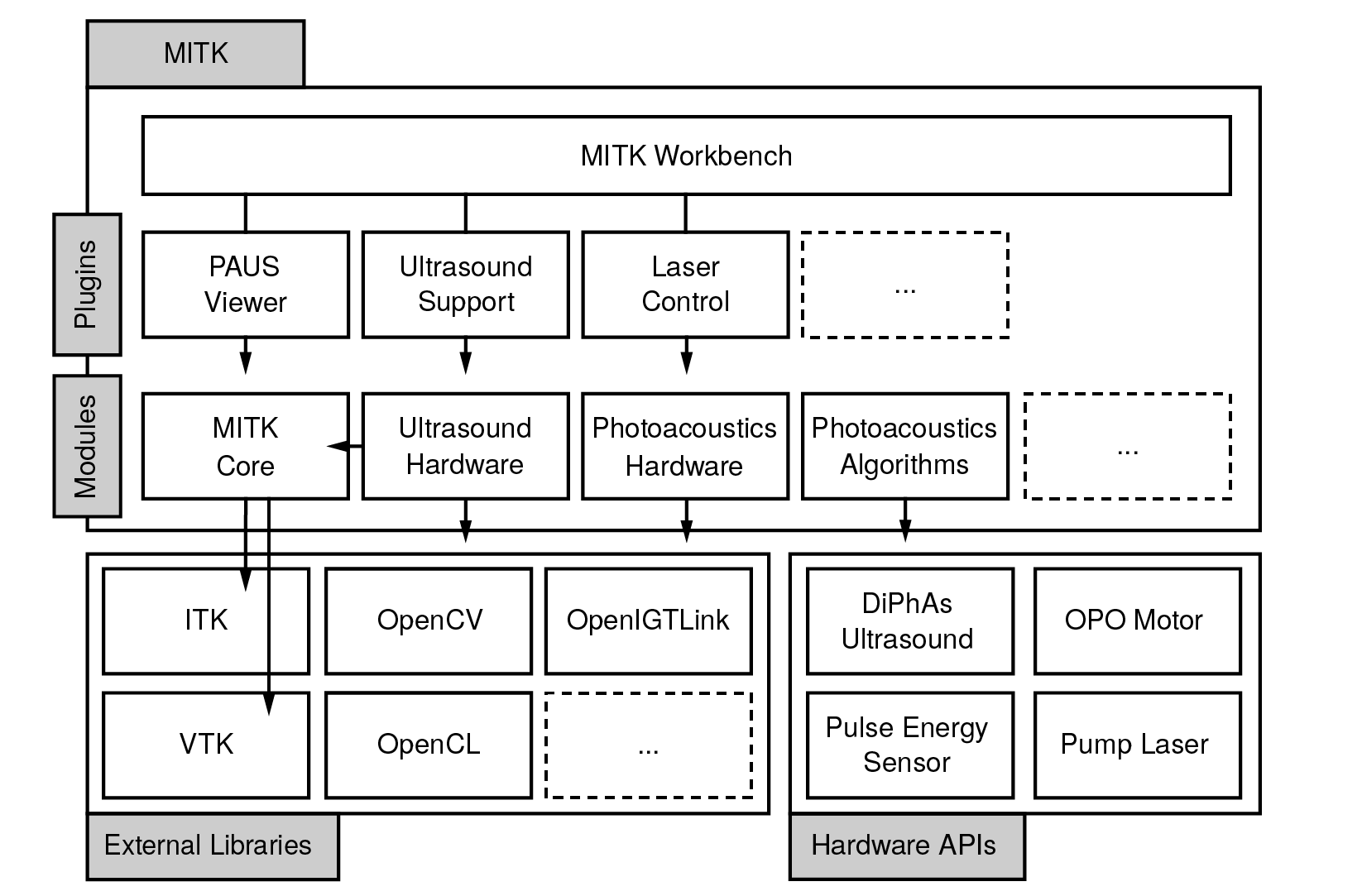}
  \end{center}
  \caption[Software components]
  {Software components of the \acf{MITK}. The MITK Workbench serves as an extendable user interface comprised of functionality provided by various \emph{Plugins}. Those Plugins also connect the user interface with functional units provided in \emph{Modules}. Those Modules can have dependencies on and draw their functionality from external libraries. The Modules also serve as an interface to external hardware and their \acf{API}s.}
  \label{fig:components} 
\end{figure}

External libraries like the Insight Toolkit (ITK)\cite{Ibanez2005} and the Visualization Toolkit (VTK)\cite{Schroeder2004} are integral dependencies of MITK. For further specialized use cases other libraries are provided through their implementations in MITK. Examples for external libraries in use are the OpenCV\cite{Culjak2012} library for motion estimation, Eigen\cite{eigenweb} for spectral unmixing and OpenCL\cite{Stone2010} for implementations of algorithms on GPU for faster execution. Data can be exchanged in real-time with systems not running via MITK or even on different PCs using MITKs implementation of OpenIGTLink\cite{Klemm2017}.
Offline processing of recorded data can be performed with the \ac{MITK} \ac{PA} image processing plugin, as was the case for the beamforming in the experiments.\cite{Kirchner2018sdmas}\footnote{using commit \texttt{https://phabricator.mitk.org/rMITK9ce68418f58b}}
Fig.\,\ref{fig:components} shows the main software components we implemented or extended for PAUS imaging. The following paragraphs describe those components in detail.

\paragraph{Hardware layer}
The hardware layer abstracts the control of the hardware used, i.e.\ the DiPhAS US system and the OPO laser. 
Hardware control has been implemented in the classes corresponding to the \emph{USHardwareDiPhAs} sub-module of the \emph{Ultrasound Hardware} module and the \emph{PhotoacousticsHardware} modules. 
The \emph{Ultrasound Hardware} module and its sub-modules rely heavily on the existing generalized architecture of MITK for US devices\cite{Marz2014}.
Communication with the DiPhAs API and therefore control over the transducer and its output is implemented in the \emph{USHardwareDiPhAs} sub-module. The transducer itself is represented as a generalized US device by a subclass of the abstract \emph{Device} class where all hardware control is centralized. The \emph{Device} class, as well as further classes, which are accessible through the  \emph{Device} class, such as an \emph{ImageSource} subclass that handles image acquisition, handle communication, storage and adjustment of necessary parameters for image acquisition, e.g.\ time gain compensation or transducer voltage, as well as the setup of the interleaved PAUS acquisition sequence.
This simple hierarchical approach allows to use a single unifying extendable interface to access multiple US devices, which can be managed in parallel. 

The \emph{ImageSource} class acquires image data from the US system through the external vendor API through a callback method. Both the raw image data, as well as images beamformed by the DiPhAS system are passed through the callback, allowing both for fast display of the beamformed US and PA images, as well as for later image processing of the raw data.
The passed image data is continuously put into a buffer which is accessed by the processing layer for later display, while the raw data can be saved directly onto a hard drive for later use. 

The \emph{PhotoacousticsHardware} module handles communication with the laser system to control the pump laser and tune the OPO. Specifically, the pump laser is controlled via serial port using MITKs \emph{SerialCommunication} class, while the OPO can be controlled using its API. The module also implements a method to allow the \emph{USHardwareDiPhAs} sub-module to read out the data from an internal pyroelectrical sensor in the laser system which acquires the current laser pulse energy. This data can be matched to the acquired PA images and used to perform corrections for laser energy fluctuations.

\paragraph{Processing layer}
Basic Processing is performed in the \emph{ImageSource} class before the acquired images are displayed. Whenever the plugin within the application layer requests a new image, a method within the \emph{ImageSource} class is called through the \emph{Device} class, which grabs the most recent image within the buffer that contains the beamformed images. 
Various filters can be applied to the image depending on the settings, such as a resampling filter (i.e.\ vertically rescaling the image), a B-Mode filter, or basic fluence corrections for pre-defined illumination geometries. Afterwards, the processed image is passed to the application layer.

\paragraph{Application layer}
Data acquisition from the DiPhAS setup is performed based on user settings specified in the MITK US Support plugin \cite{Marz2014} which has been extended to work with our custom \ac{DAQ}\footnote{The data acquisition was performed with commit \texttt{https://phabricator.mitk.org/rMITK2d2ebd4f22fa}}. The US Support plugin communicates with the USHardwareDiPhAS sub-module through an instance of its \emph{Device} subclass, which it accesses through MITK's micro service functionality; through micro services, a module is able to register as a specific service, which can be then requested by other modules or plugins.
Various PA specific image acquisition settings are available within the US Support plugin. These preferences are handed over to the \emph{USHardwareDiPhAS} sub-module which is responsible for communication with the hardware \ac{API}. The plugin also passes image data acquired through the device's \emph{ImageSource} class to the \emph{PAUSViewer} plugin. The \emph{PAUSViewer} plugin has been implemented specifically for the use case of PAUS imaging and provides a view that presents corresponding images of both modalities side by side in real-time, with options to set specific level windows and various colormaps. To set the triggers for the Laser system, define fast tuning wavelength sequences, as well as to control the status of the laser, the \emph{Laser Control} Plugin was implemented. The plugin serves as a user interface for the \emph{PhotoacousticsHardware} module. The application layer in general is highly customizable and extendable depending on the use case or specific problem at hand.

\section{Clinical Sample Application}

An example for a possible clinical application of \ac{PA} imaging is the visualization of transmural inflammatory processes such as large cell arteritis, i.e. giant cell arteritis\cite{Hunder1990}, where \ac{PA} imaging could be used in the diagnostic workup. As an initial step towards this use-case we image the carotid artery of a healthy human volunteer and estimate blood oxygenation as a qualitative validation. The long term goal for \ac{PA} imaging of the carotid artery is to investigate if and how multispectral \ac{PA} imaging can be used in the diagnosis of inflammatory processes in arteries. In addition, \ac{PA} imaging may be useful in the evaluation of plaque morphology in carotid artery stenosis or in the planning phase of surgeries for head and neck cancers invading the carotid artery.

One of the main issues during the sequential measurement of multiple multispectral \ac{PA} images of the carotid artery is inter-frame motion, which leads to invalid results due to intense motion artifacts. If the motion is periodic, these artifacts might be somewhat mitigated by frame averaging \cite{Kim2013} and there exist approaches in \ac{PA} computed tomography which assume rigid motion \cite{Willemink2006} or perform gated image acquisition \cite{Xia2014} to minimize motion artifacts. But frame averaging in moving structures will always improve imaging results at the expense of resolution and imaging time as well as temporal resolution. Also, the motion of the carotid artery and the surrounding tissue is not rigid. Furthermore, gating is impractical due to the need for additional equipment and the about 20-fold decrease in imaging frame rate to approximately 1\,Hz which in turn leads to increased motion artifacts due to the movement of patient and physician especially when using free hand probes. The following section shows how we aim to demonstrate the applicability of our framework to \emph{in vivo} blood oxygenation estimation.

\subsection{Motion-compensated blood oxygenation estimation}
In the following we present the methods used for our platform demonstration experiments: (1) The spectral unmixing method for $\mathrm{sO_2}$ estimation and (2) our new inter-frame motion correction approach. Both methods are implemented as python extensions with minimal overhead and can be used via the python interface offered by MITK.\footnote{see \texttt{http://docs.mitk.org/nightly/mitkPython\_Overview.html}}.

\subsubsection{Blood oxygenation estimation with spectral unmixing}
\label{sec:so2}
$\mathrm{sO_2}$ estimation with spectral unmixing from PA images requires a number of acquired wavelengths of at least the number of unmixed chromophores (two). Using more wavelengths will make the estimation more robust.\cite{Tzoumas2016}
We record a \emph{sequence} of raw \ac{PA} data at five wavelengths in the near infrared. Based on Luke et al.\cite{Luke2013} and considering the power spectrum of our laser source we skew our wavelength sequence from an equidistant spacing towards wavelengths where the differences between absorption of oxygenated and deoxygenated hemoglobin are most significant. The acquisition wavelength sequence is measured by a spectrometer (HR2000+, Ocean Optics, Dunedin, USA) to account for errors in OPO calibration.

Spectral unmixing is performed using a non negative constrained linear least squares solver\cite{Lawson1995}\footnote{\texttt{scipy.optimize.nnls} using python 2.7} on the sets of five B-Mode images. The unmixing results for oxygenated (HbO$_2$) and deoxygenated hemoglobin (Hb) are used to calculate blood oxygenation and total hemoglobin (THb):
\begin{equation}
    \mathrm{sO_2} = \frac{\mathrm{HbO_2}}{\mathrm{Hb + HbO_2}}, \qquad \mathrm{THb} = \mathrm{Hb + HbO_2}.
\end{equation}\\
$\mathrm{sO_2}$ is visualized by masking the results for low values of $\mathrm{THb}$.

\subsubsection{PAUS inter-frame motion correction}
\label{sec:of}
We propose a method which uses \ac{US} images to compensate for intra-sequence motion of the \ac{PA} images.
Our PAUS imaging system acquires interleaved \ac{US} images with minimal delay after each \ac{PA} image. We estimate the \emph{optical flow}\cite{Farneback2003} of each \ac{US} image in a sequence with respect to the first \ac{US} image in that sequence. We then use these estimated optical flows to warp their corresponding \ac{PA} images. As ``Optical flow is the distribution of apparent velocities of movement of brightness patterns in an image''\cite{Horn1981}, it is necessary to estimate the optical flow in the \ac{US} images instead of the \ac{PA} images, as brightness and the patterns having that brightness vary strongly in a multispectral \ac{PA} acquisition sequence.

The specific optical flow implementation we use is by Farnebäck et al.\cite{Farneback2003}\,and part of the Open CV library\footnote{as \texttt{cv2.calcOpticalFlowFarneback} in phython 2.7}. Knowing that the motion we want to compensate for is relatively small and quite homogeneous with no small structures moving independent of the surrounding tissue, we want an optical flow estimation which is approximated with a smooth surface -- a blurred motion field. In addition, we aim for a fast (real-time capable) computation. Because of that we only perform two iterations (\texttt{iterations} = 2) and chose a large averaging window and neighborhood (specifically: \texttt{winsize} = 40, \texttt{poly\_n} = 7, \texttt{poly\_sigma} = 1.5). This will result in a blurred motion field that has the desirable side-effect of yielding a more robust algorithm.

\subsection{Validation experiments}
The purpose of our validation experiments was to demonstrate the applicability of our platform to a translational research field.
The validation experiments of the MITK PAUS real-time imaging platform were performed on data acquired with the platform. Two experiments were performed with the goal to (1) estimate blood oxygenation in peripheral vessels, namely the radial and ulnar arteries and to (2) estimate blood oxygenation the carotid artery. Both (1) and (2) require oxygenation estimation as detailed in section\,\ref{sec:so2} and (2) requires additional motion correction as detailed in section\,\ref{sec:of}. While the data acquired for the experiments was processed and visualized as B-Mode images (both PA and US) in real-time, we performed the analysis for the experiments offline on the data recorded beforehand by MITK. However, oxygenation estimation and motion correction have been both implemented to be real-time capable. Both \emph{in vivo} PAUS experiments were performed free hand on healthy human volunteers while aiming to hold the probe as still as possible while acquiring approximately twenty seconds of data. Imaging of the carotid artery was performed by a vascular surgeon.
The acquired raw \ac{PA} data was corrected for fluctuations in laser pulse energy and then beamformed using the commonly used \ac{DAS} algorithm\cite{Kim2016} with Hanning apodization.

\section{Results}
During the acquisition of the presented data, the user had a real-time view of both US and PA images at frame rates of 13 to 20\,Hz.
The fast tuned \ac{PA} acquisition wavelength sequence of the OPO laser was measured as $(722, 756, 831 , 907, 943)$\,nm with an accuracy of 1.5\,nm. In the following we show oxygenation measurements in peripheral vessels and the effect of motion correction on $\mathrm{sO_2}$ estimations in a carotid artery.

\subsection{Blood oxygenation in peripheral vessels}
\begin{figure}[h!bt]
  \begin{center}
  \includegraphics{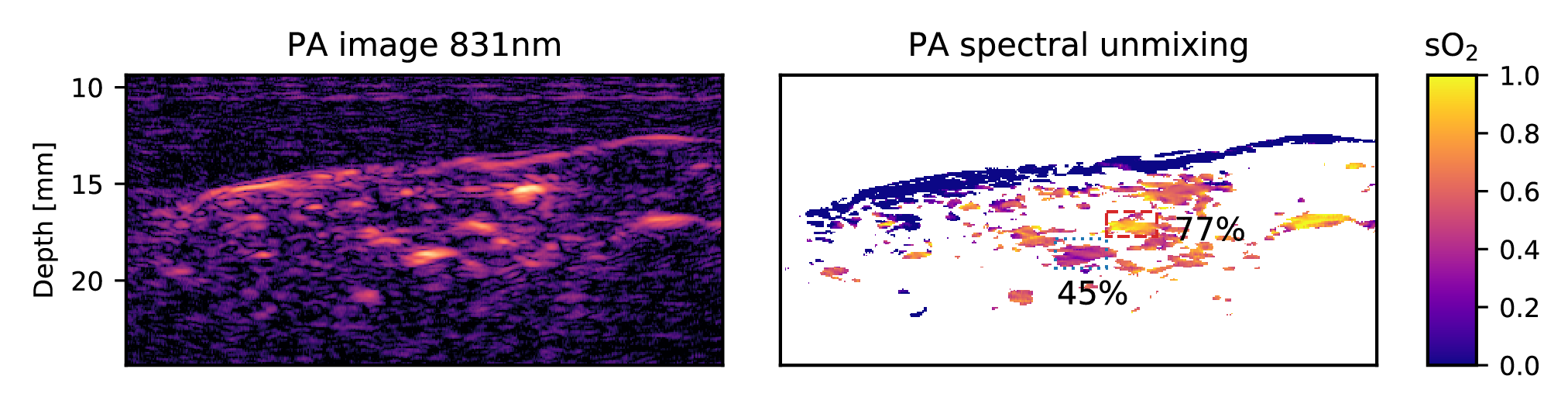}
  \end{center}
  \caption[example]
  {Representative \acf{PA} B-mode image and blood oxygenation ($\mathrm{sO_2}$) estimation in the radial artery and accompanying vein of a healthy human volunteer. The \ac{PA} image is shows logarithmically compressed. The $\mathrm{sO_2}$ estimation is visualized for pixels with relevant total hemoglobin results unmixed from one sequence of five wavelengths. The dashed red box deli the region of interest for arterial sO$_2$, the dotted blue box for venous. The average sO$_2$ values for this image are denoted on the boxes.}
  \label{fig:radial:ulnar} 
\end{figure}

Fig.\,\ref{fig:radial:ulnar} shows a representative example of $\mathrm{sO_2}$ estimation in the radial artery and accompanying vein. $\mathrm{sO_2}$ is visualized by thresholding the total hemoglobin from the unmixing results. The average $\mathrm{sO_2}$ in the radial and ulnar artery was $\mathrm{SaO_2} = 72\,\% \pm 7\,\%$ averaged over a total of $n = 436$ mean oxygenations in a region of interest as marked in Fig.\,\ref{fig:radial:ulnar}. The radial artery has been scanned twice, likewise the ulnar artery. The $\mathrm{sO_2}$ in the accompanying vein of the radial artery was on average $\mathrm{SvO_2} = 38\,\% \pm 9\,\%$ over a total of $n = 238$ mean oxygenation estimations during two scans on the same healthy human volunteer.\\

\subsection{Motion correction and blood oxygenation estimation in the carotid artery}
\begin{figure}[hbt!]
  \begin{center}
  \includegraphics{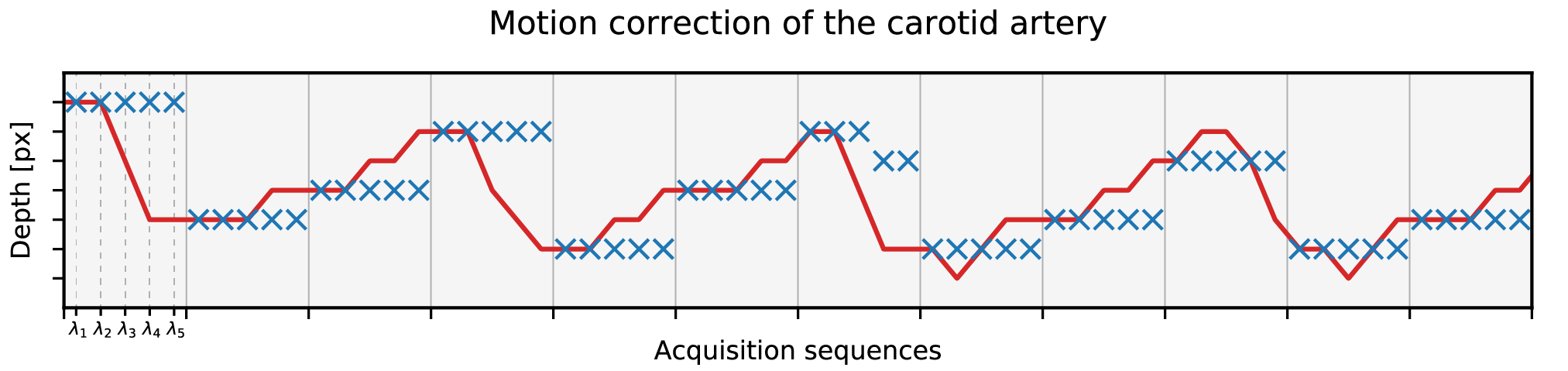}
  \end{center}
  \caption[example]
  {Illustration of motion correction in the first acquisition sequences; each sequence consisting of five images acquired at different wavelengths $\lambda$. The depth ($y$ position of the upper arterial wall) in the \acf{US} images is plotted before (red, $-$) and after the application of optical flow motion compensation (blue, $\times$). The periodic shift in position due to expansion of the carotid artery at a frequency corresponding to the resting heart rate of the volunteer is apparent.}
  \label{fig:flow}
\end{figure}

Estimating the optical flow for four $(350\times200$\,px sized) US images of a sequence relative to the first US images in that sequence and warping the acquired PA and US images accordingly took $(120\pm10)$\,ms (averaged over $n = 84$ sequences acquired in one scan) with python on a single core (2.6\,GHz Intel Core i5-3230M). Fig.\,\ref{fig:flow} illustrates the $y$ position of the upper arterial wall in the US images of the acquisition sequences before and after the application of our optical flow based method by plotting the position of the maximum intensity pixel in the center of the upper arterial wall.

\begin{figure}[hbt!]
  \begin{center}
  \includegraphics{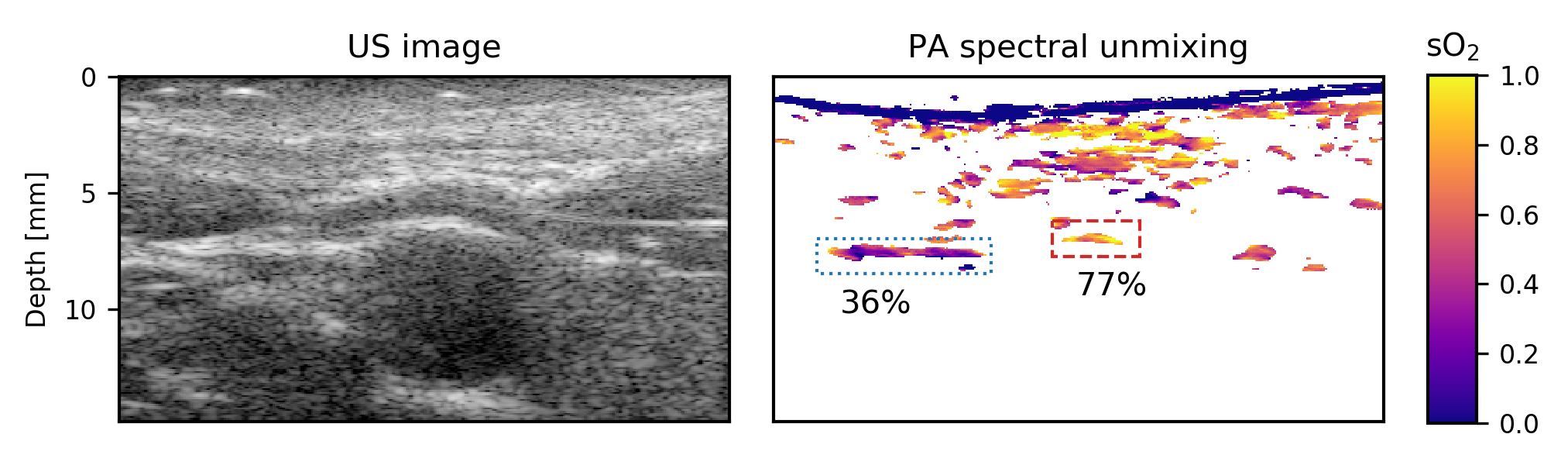}
  \end{center}
  \caption[example]
  {Representative \acf{US} image of a carotid artery with corresponding blood oxygenation (sO$_2$) estimation with applied \emph{optical flow} motion correction. The dashed red box is the region of interest for arterial blood oxygenation(sO$_2$), the dotted blue box for venous. The average sO$_2$ values for this image are denoted on the boxes.
  }
  \label{fig:carotid} 
\end{figure}

\begin{figure}[hbt!]
  \begin{center}
  \includegraphics{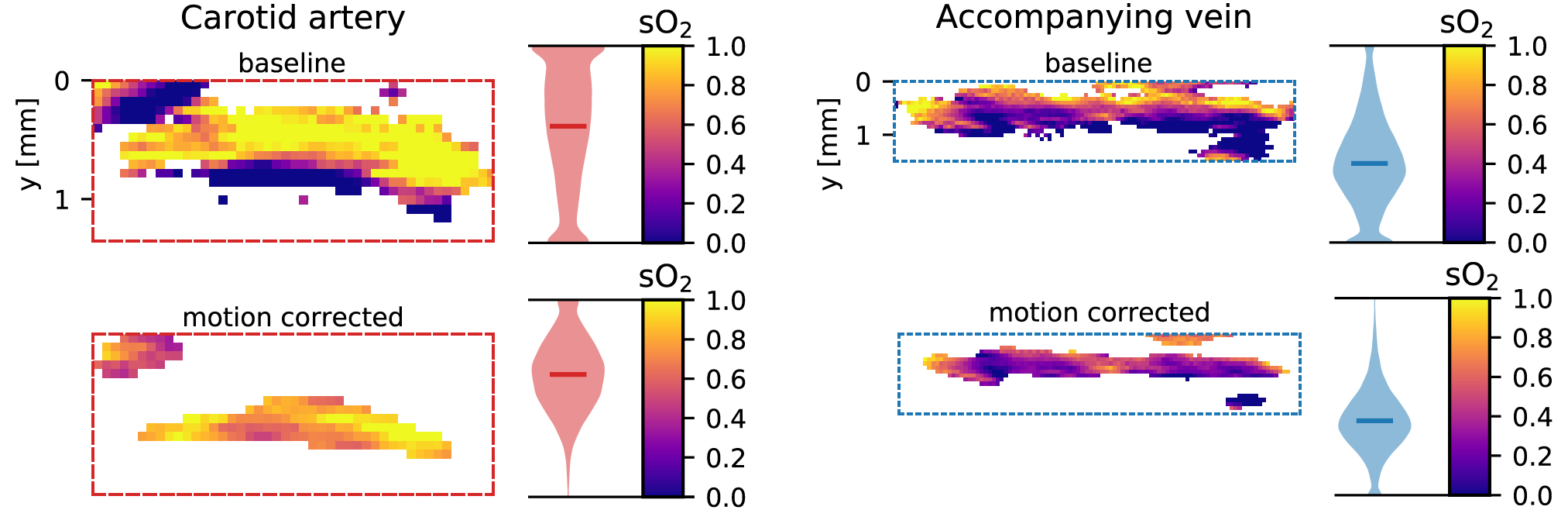}
  \end{center}
  \caption[example]
  {Representative blood oxygenation (sO$_2$) estimation by unmixing from \acf{PA} images of a carotid artery (left) and vein (right) as well as the same estimations with applied \emph{optical flow in sequence (ofis)} motion correction (bottom). The motion artifact in the artery is apparent in the top-left image and is vastly suppressed in the bottom-left. The attached violin plots show the distribution of the sO$_2$ measurements over all pixels in the region of interest from $n = 84$ sequences. A multimodal distribution of sO$_2$ in the carotid artery is apparent without motion correction. The inter quartial ranges (IQR) for the plotted distributions are: in arterial blood a reduction from IQR(SaO$_2) = 50\,\% $ to IQR(SaO$_2^{ofis}) = 25\,\% $ with motion correction; in venous blood a reduction from IQR(SvO$_2) = 29\,\% $ to IQR(SvO$_2^{ofis}) = 19\,\% $.
  }
  \label{fig:artifacts}
\end{figure}

Fig.\,\ref{fig:carotid} shows a section of the carotid artery and its corresponding oxygenation. The oxygenation was estimated based on the motion corrected stack of \ac{PA} images. Average sO$_2$ in carotid artery and jugular vein is denoted on the regions of interest shown as red and blue boxes. An example of artifacts resulting from spectral unmixing on sequences not corrected for motion is shown in Fig.\,\ref{fig:artifacts}.
We measured the oxygenation saturation with $84$ image sequences of the carotid artery. Estimating on images uncorrected for motion yielded mean(SaO$_2) = $ 59\,\% when averaging over the $n = 29,861$ pixels over the THb threshold. The standard deviation of the measurement was std(SaO$_2) = 31\,\%$. After \emph{optical flow in sequence (ofis)} motion correction we measured SaO$_2^{ofis} = (63 \pm 17)\,\%$ over $n = 12,172$ pixels. In the accompanying jugular vein the estimation changed from SvO$_2 = (40 \pm 21)\,\%$ without motion correction to SvO$_2^{ofis} = (38 \pm 15)\,\%$ with motion correction. The distributions of sO$_2$ over all pixels is shown in the violin plots of Fig.\,\ref{fig:artifacts}. As some of these are no normal distributions, we also determined their inter quartial ranges (IQR): For arterial blood a reduction from IQR(SaO$_2) = 50\,\% $ to IQR(SaO$_2^{ofis}) = 25\,\% $ for venous blood a reduction from IQR(SvO$_2) = 29\,\% $ to IQR(SvO$_2^{ofis}) = 19\,\% $.

%quartiles 1,2,3
%before vein 25.2 39.2 54.3
%after vein 27.6 36.9 47.0

%before artery 36.2 64.0 86.6
%after artery 49.7 62.3 75.0

\section{DISCUSSION}

In this paper we present a real-time PAUS imaging platform with a linear array probe using a fast tuning laser. Within the open-source software platform \ac{MITK} we provide implementations for direct control of the US DAQ and laser systems and their components. The presented software components are highly reusable and extensible due to the modular architecture of {MITK}. This should make it possible to use the project as a starting point for translational research projects with PA imaging. For each specific laser or DAQ system, only the hardware layer would have to be extended to comply with new APIs.

Furthermore we provide a method for correcting inter-frame motion as encountered in a clinical sample application which we presented as a demonstration use case of our system. In this clinical sample application we imaged the carotid artery of a healthy human volunteer. In this context we were faced with problematic inter-frame motion due to the pulsing artery. To address this issue we presented a new method for motion correction of multispectral \ac{PA} image sequences using \emph{optical flow} in seqence (ofis) on corresponding \ac{US} images. Run time measurements of this methods show the real-time capability of the method considering motion correction of a sequence of five corresponding \ac{PA} images is performed in 120\,ms in our python implementation running single threaded on a consumer CPU core -- while the sequence acquisition takes 250\,ms with out fast tuning OPO. The motion correction method succeeds in reducing motion artifacts and the sO$_2$ estimations with applied motion correction make it possible to clearly distinguish between arterial and venous blood. The variation of estimated sO$_2$ is reduced as shown by the drop in standard deviation when correction for motion and illustrated in Fig.\,\ref{fig:artifacts}. This shows a higher precision of sO$_2$ estimation by spectral unmixing when correcting for inter-frame motion. It is to our knowledge the first inter-frame motion correction approach using corresponding \ac{US} images or optical flow. 

While the sO$_2$ estimates when using motion correction are more consistent and closer to the physiological values in a healthy human\cite{Zander1990}, there is a systematic underestimation of oxygenation with our data. This is also the case in more superficial vessels, but to a lesser degree. We attribute this underestimation to (1) fluence effects due to a high overall oxygenation in tissue -- this should be addressed quantitatively\cite{Tzoumas2016, Kirchner2018} and to a lesser degree (2) noise levels.

We have presented a starting point for translational PAUS imaging research integrated in the open-source MITK platform. We validated its performance on a clinical sample application, were we were able to show that we can image the carotid artery multispectrally by correcting for inter-frame motion using an optical flow based approach.

\appendix
\acknowledgments            
The authors would like to acknowledge support from the European Union through the ERC starting grant {\footnotesize COMBIOSCOPY} under the New Horizon Framework Programme under grant agreement ERC-2015-StG-37960.
We also would like to thank the MITK team for providing the open-source development and testing infrastructure that was used in this project, as well as E.~Stenau for fruitful discussions of motion correction.

\end{document}